\documentstyle[prl,floats,aps,epsf]{revtex}
\begin{document}
\advance\textheight by 0.5in
\advance\topmargin by -0.25in
\draft

\twocolumn[\hsize\textwidth\columnwidth\hsize\csname@twocolumnfalse%
\endcsname

\title{ 
Quantum Brownian Motion in a Periodic Potential\\
and the Multi Channel Kondo Problem}
 
\author{Hangmo Yi and C.L. Kane}
 
\address{Department of Physics and Astronomy,
 University of Pennsylvania \\
 Philadelphia, Pennsylvania 19104}

\date{February 19, 1996}

\maketitle

\begin{abstract}

We study the motion of a particle in a 
periodic potential with Ohmic dissipation.
In $D=1$ dimension it is well known that there are two phases depending
on the dissipation:
a localized phase with zero temperature mobility $\mu=0$ and a fully
coherent phase with $\mu$ unaffected by the periodic potential.
For $D>1$, we find that this is also the case for a 
Bravais lattice.  However, for non symmorphic lattices, such
as the honeycomb lattice and its $D$ dimensional
generalization, there is a new intermediate phase with a universal
mobility $\mu^*$.  We study this intermediate fixed point
in perturbatively accessible regimes.  In addition, we
relate this model to the Toulouse limit of the 
$D+1$ channel Kondo problem.
This mapping allows us to compute $\mu^*$ exactly
using results known from conformal field theory.   
Experimental implications are discussed for 
resonant tunneling in strongly coupled Coulomb blockade
structures and for multi channel Luttinger liquids.

\end{abstract}
\pacs{PACS numbers: 05.40+j, 05.30.-d, 72.15Qm, 73.40Gk}
\vskip -0.5 truein
]

The quantum mechanics of a particle in a periodic 
potential coupled to a dissipative environment is a 
fundamental problem in condensed matter physics\cite{caldeira}.
A simple theory based on the 
Caldeira-Leggett model of Ohmic dissipation was 
proposed in the mid 1980's as a possible description of
the motion of a heavy charged particle in a metal\cite{fisher}.
In a one dimensional periodic potential it was shown
that there are two zero temperature ($T=0$) phases. 
For weak friction,
the particle diffuses freely as if the periodic
potential were absent.  When the friction exceeds
a critical value, however, the particle is localized 
in one of the minima of the potential.

Recently there has been renewed interest 
in this quantum Brownian motion (QBM) model in connection with quantum impurity 
problems\cite{kane} and boundary conformal field theory\cite{ludwig}.  
It is isomorphic to the problem of tunneling through a 
barrier in a Luttinger liquid, 
which is relevant to experiments in quantum wires\cite{Tarucha} and 
tunneling in quantum Hall edge states\cite{Webb}.
Here the ``coordinate" of the ``particle" 
is the number of electrons that tunnel past the barrier.
The periodic potential arises from the discreteness
of the electron's charge.  The Luttinger liquid's modes 
play the role of the dissipative bath.
The particle's mobility corresponds to the 
electrical conductance.

There are often multiple electron channels, due
to spin and transverse degrees of freedom.
The impurity problem then maps to a multi-dimensional
periodic potential.  In addition to the extended
and localized phases, in two dimensions it has 
been shown that there can be additional non trivial 
phases\cite{kane2,callan}, 
which may be accessed by tuning to a resonance.   
Using a similar analysis, Furusaki and Matveev
recently found a similar intermediate phase
in a model of resonant tunneling through a Coulomb
blockade structure\cite{furusaki}.  They argued that the resonance 
fixed point is that of the multi channel Kondo problem.

In this paper we consider the general problem of QBM
on periodic lattices.  We show that
the lattice symmetry plays a crucial role in 
determining the $T=0$ phases.  
For the honeycomb lattice and
its $D$ dimensional generalization, there is a $T=0$ phase
described by an intermediate fixed point, which we
relate to the $D+1$ channel Kondo fixed point. 
Exploiting the mapping onto the Kondo problem, we 
compute exactly the fixed point mobility and critical
exponents by borrowing results from conformal field theory.

Integrating out the bath degrees of freedom, 
the QBM model is described by the 
Euclidean action\cite{caldeira,kane}
\begin{equation}
S = S_0[{\bf r}(\tau)] - \int {d\tau\over\tau_c} 
\sum_{\bf G} v_{\bf G} e^{i 2\pi {\bf G}\cdot{\bf r}(\tau)},
\end{equation}
where ${\bf r}$ is the coordinate of the particle
and $v_{\bf G}$ are
dimensionless Fourier components of the periodic potential,
defined at the reciprocal lattice vectors ${\bf G}$.
(${\bf G}$ is defined so that 
${\bf G}\cdot{\bf R}$ is an integer
for any lattice vector ${\bf R}$.)
The coupling to the dissipative bath is described by
\begin{equation}
S_0[{\bf r}(\tau)] = 
{1\over 2} \int d\omega |\omega| e^{|\omega| \tau_c}
|{\bf r}(\omega)|^2,
\end{equation}
where $\tau_c$ is a short time cutoff.
The friction is proportional
to the coefficient of this term.  However, by rescaling ${\bf r}$ and 
$\bf G$ this coefficient may be fixed.   
The lattice constant thus controls the strength of the friction.    
In accordance with Ref. \onlinecite{kane}, we define the dimensionless
parameter $g = |{\bf R}_{\rm min}|^{-2}$, where $|{\bf R}_{\rm min}|$
is the Bravais lattice constant.  $g$ is
inversely proportional to the friction.
A $1+1$ dimensional version of this theory has recently been 
analyzed by Kondev and Henley\cite{henley}.

Our system may be characterized by the mobility,
which describes the average velocity of the particle in response to
a uniform applied force.
We define the dimensionless mobility $\mu$ as the
ratio of the mobility to the ``perfect" mobility
obtained in the absence of the periodic potential.
When $v_{\bf G} =0$, $\mu = 1$.
$\mu$ may be computed from linear response theory,
\begin{equation}
\mu =  (2\pi/D) \lim_{\omega\rightarrow 0} |\omega| 
 \langle |{\bf r}(\omega)|^2 \rangle .
\label{eq:muGen}
\end{equation}

The effect of the periodic potential may be analyzed perturbatively
in either of two limits.  
A weak potential may be studied by
considering the renormalization group (RG) 
flows to leading order in $v_{\bf G}$,
\begin{equation}
dv_{\bf G}/d\ell = (1 -  |{\bf G}|^2 ) v_{\bf G}. 
\end{equation}
Clearly, if the shortest reciprocal lattice vector
satisfies $|{\bf G}_{\rm min}| > 1$, then all
$v_{\bf G}$ are irrelevant.  The ``small barrier" limit
in which the particle diffuses freely is thus
perturbatively stable.  On the other hand, if
$|{\bf G_{\rm min}}| <1$, then the
system flows to a different strong coupling phase.

When the barriers are large, the particle 
is localized in one of the minima of the potential
with a small probability for tunneling to another.  
It is then more natural to consider a dual representation
in which the the partition function is expanded
in powers of the ``fugacity" of these tunneling events\cite{fisher,kane}.
For a Bravais lattice, this may be generated by expanding the dual action,
\begin{equation}
S = S_0[{\bf k}(\tau)] - \int {d\tau\over\tau_c}  
\sum_{\bf R} t_{\bf R} e^{i 2\pi {\bf R}\cdot{\bf k}(\tau)}.
\end{equation}
$t_{\bf R}$ may be interpreted as the matrix element for the particle
to tunnel between minima connected by a lattice vector ${\bf R}$.
Equivalently, ${\bf k}(\tau)$ describes the particle's trajectory in 
momentum space in a potential
with the symmetry of the reciprocal lattice.
The RG flows to leading order in $t_{\bf R}$ are then
\begin{equation}
dt_{\bf R}/d\ell = (1- |{\bf R}|^2) t_{\bf R}. 
\label{eq:dtdl}
\end{equation}
The ``large barrier" phase is thus 
perturbatively stable provided the
shortest lattice vector satisfies $|{\bf R}_{\rm min}| > 1$.

For a one dimensional lattice, $|{\bf R}_{\rm min}|^2 = 1/g$
and $|{\bf G}_{\rm min}|^2 = g$.
Thus either the small or the large barrier limit
is stable, but not both.  There are two phases:  for
$g<1$ the system is localized and for $g>1$ the system has perfect
mobility.  Clearly, this is also the case in higher 
dimensions for a lattice with cubic symmetry.   

In contrast, for a triangular lattice,
$|{\bf R}_{\rm min}|^2 = 1/g$, but  $|{\bf G}_{\rm min}|^2 = 4g/3$.
It follows that for $3/4 < g < 1$, {\it both} the small and large barrier
limits are stable.  There must therefore be
an unstable fixed point separating
the two phases, as indicated in Fig. 1a.  A similar intermediate fixed point
occurs in the single barrier problem of a spin 1/2 Luttinger
liquid\cite{kane2}.  

A perturbative analysis of this fixed point is
possible for small $v$ in the vicinity of $g = 3/4$.
Specifically, consider a model with $v_{\bf G} = v$
for the six ``nearest neighbor" reciprocal lattice vectors.
For $v>0$, this produces a potential with minima forming
a triangular lattice.  
For $g = 3/4(1+\epsilon)$, the RG flow 
to second order in $v$ is \cite{kane2}
\begin{equation}
dv/ d\ell =  -\epsilon v + 2 v^2.
\end{equation}
Provided $v>0$ and $\epsilon>0$, 
there is an unstable fixed point $v^* = \epsilon/2$,
with RG eigenvalue $\epsilon$.  The dimensionless
mobility at this fixed point is universal, 
$\mu^* = 1 - (3\pi^2/ 2) \epsilon^2 $.

\begin{figure}
\epsfxsize=2.8in
\centerline{ 
\epsffile{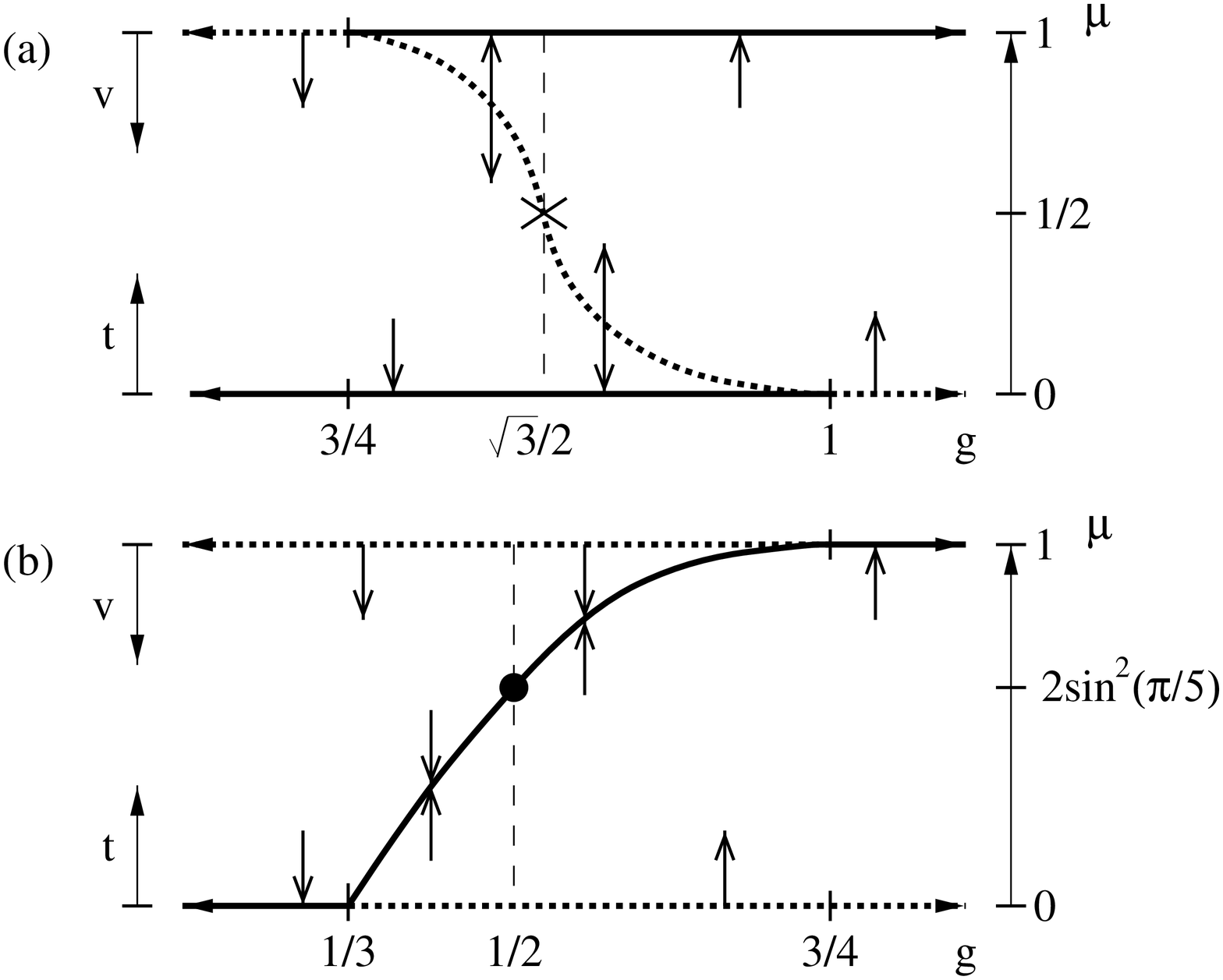} 
}
\caption{Flow diagrams for the (a) triangular
lattice and (b) honeycomb lattice.  
The top (bottom) lines represent the small $v$ ($t$) limits.
Stable (unstable) fixed points
are depicted by solid (dotted) lines, and
arrows indicate the RG flows.  The fixed point
mobility is known perturbatively near 
$g = 1/3, 3/4, 1$
and at the points indicated by the dashed lines.
}
\end{figure}

A similar analysis is possible in the dual theory for small $t$.
Since the dual lattice is also triangular, the
results are identical, given the substitutions
${\bf r} \rightarrow {\bf k}$,
${\bf G} \rightarrow {\bf R}$,
$g \rightarrow 3/(4g)$
and $ v\rightarrow t$.
$\mu$ is mapped to $1 - \mu$.
For $g=1-\epsilon$, the fixed point is at 
$ t^*= \epsilon/2$, with an exponent $\epsilon$, and
mobility $\mu^* = (3\pi^2/ 2) \epsilon^2 $.
For $g = \sqrt{3}/2$, the theory is self dual,
which implies that the fixed point mobility is $\mu^* = 1/2$.
Piecing these results together, we obtain the flow diagram in Fig. 1a.

When $v<0$, the minima of the potential described
above form a honeycomb lattice.  
The honeycomb lattice is equivalent to the triangular lattice
described above with a two site basis.
If the triangular lattice constant is $1/\sqrt{g}$,
then we again have
$|{\bf G}_{\rm min}|^2 = 4g/3$.  
However $|{\bf R}_{\rm min}|^2 = 1/(3g)$ 
is now shorter. 
Thus, for $1/3 < g < 3/4$ both the large and small
barrier limits are {\em unstable}, so that there must
be a stable fixed point describing a new
intermediate coupling phase.    
With $v$ negative, Eq. (1) may also be viewed as the tight binding 
representation of a triangular lattice with $\pi$ flux
per plaquette, whose reciprocal is the honeycomb lattice.
Our analysis applies to this dual theory as well.  
Callan et. al. \cite{callan} have recently found
similar intermediate phases on a square lattice
with magnetic flux.

A perturbative analysis is again possible in the large and small
barrier limits.  For small barriers the 
fixed point of Eq. (7) is stable for $v<0$ and $\epsilon<0$.
The RG eigenvalue and mobility at the fixed point are the same
as above.  In the large barrier theory we must
keep track of the two site basis of the honeycomb lattice.
There are three nearest neighbors ${\bf R}$ for each site on
the A sublattice.  For the B sublattice the nearest neighbors
are $-{\bf R}$.  The tunneling must alternate between the sublattices.
This can be incorporated in the dual theory by 
introducing a spin $1/2$ degree of freedom.
For nearest neighbor hopping the dual action is then 
\begin{equation}
S = S_0[{\bf k}]
-\int {d\tau\over\tau_c} \sum_{\bf R}
 t \left[\tau^+ e^{i 2\pi {\bf R}\cdot{\bf k}} + 
       \tau^- e^{-i 2\pi {\bf R}\cdot{\bf k}}\right],
 \label{eq:Sthoneycomb}
\end{equation}
where ${\bf R}$ are among the 3 nearest neighbor lattice vectors of
sublattice A and $\tau^\pm$ are spin 1/2 operators, $\sigma^\pm/2$.
The intermediate fixed point may now be accessed perturbatively
for $g = (1/3)(1+\epsilon)$.   
We have computed the RG flow equation to order $t^3$,
\begin{equation}
dt/d\ell = \epsilon t -  3 t^3.
\end{equation}
For $\epsilon>0$ there is a stable fixed 
point at $t^* = \sqrt{\epsilon/3}$, with RG 
eigenvalue $2\epsilon$.  The fixed point mobility is
$\mu^* = \pi^2 \epsilon$.

The flow diagram for the honeycomb lattice as
a function of $g$ is summarized in Fig. 1b.  
Unlike the cubic and triangular Bravais lattices,
the $T=0$ mobility does not exhibit
a discontinuous jump from $0$ to $1$ as $g$ is
increased.  Rather, the mobility interpolates
smoothly between the two limits in the 
intermediate phase for $1/3<g<3/4$.
Below we show that for $g=1/2$, the intermediate fixed point
is that of the 3 channel Kondo problem.  Exploiting 
this mapping, the mobility for $g=1/2$ may be computed exactly.
The generalization of the above analysis to other
lattices and higher dimensions is straightforward
and will be given elsewhere.  In general, the 
existence of a stable intermediate phase requires
a non symmorphic lattice symmetry, with a vector
connecting equivalent sites that is shorter than 
any lattice translation. 

We now relate the stable intermediate fixed point
to the multi channel Kondo problem by 
identifying the lattice symmetry in the Kondo problem. 
The Hamiltonian of the anisotropic $N$ channel Kondo model is\cite{andreas},
\[
{\cal H} = i v_F \sum_{a,s} \int dx \psi_{as}^\dagger\partial_x\psi_{as}
+ 2\pi v_F \sum_{i,a} J_i S_{\rm imp}^i s^i_a(0),
\]
where $a,s,i$ are channel, spin, and space indices, 
$S_{\rm imp}^i$ is the impurity spin,
and $s^i_a(0) = \psi_{as}^\dagger(0) (\sigma^i_{ss'}/2) \psi_{as'}(0)$
is the electronic spin in channel $a$ at $x=0$.  
We consider an anisotropic model, characterized by dimensionless
couplings $J_z$ and $J_x = J_y = J_\perp$.
Our analysis closely parallels that of 
Emery and Kivelson for the two channel Kondo problem\cite{emery}. 
We first bosonize the theory, and then do a rotation
in spin space which transforms the $J_z$ interaction.
Upon integrating out the degrees
of freedom away from $x=0$, we obtain a 
theory in terms of the boson fields at the impurity
which closely resembles the lattice models studied in this
paper.  The details of this mapping will be 
presented in a longer article, however its essence 
may be understood quite simply.
\begin{figure}
\epsfxsize=2.0in
\centerline{
\epsffile{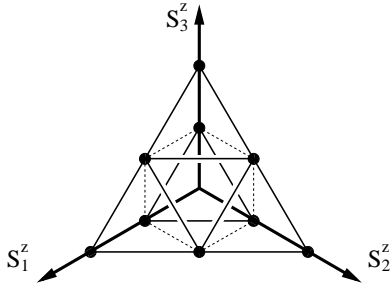}
}
\caption{Lattice of spin states for the 3 channel Kondo problem,
which form two planes with constant $S_1^z+S_2^z+S_3^z$.}
\end{figure}

When $J_\perp = 0$ the states
of the system may be characterized by the total spin 
$S_a^z$ in each of the $N$ channels.   The possible 
values of $S_a^z$ form a $N$ dimensional cubic lattice.  
$J_\perp$ ``hops" the system between
sites on this lattice.  Since $\cal H$ conserves
the total spin of the electrons plus the impurity,
the system is constrained to lie on one of two 
lattice planes with constant $S_{\rm imp}^z + \sum_a S_a^z$
where $S_{\rm imp}^z = \pm 1/2$.
For $N=3$ each lattice plane forms a triangular
lattice, as sketched in Fig. 2.  
Viewed from the (111) direction, the two lattice planes 
form a ``corrugated" honeycomb lattice in which the two triangular
sublattices are displaced in the perpendicular direction.  
For general $N$, the ``lattice planes"
consist of two interpenetrating $N-1$ dimensional close packed
lattices.  For $N=4$, they form a corrugated diamond lattice. 

Now consider QBM on such a lattice described by  
\[
S = S_0[{\bf k}]
- \int {d\tau\over\tau_c} \sum_{{\bf R}_\parallel} 
t\left[\tau^+ e^{i 2\pi({\bf R_\parallel}\cdot 
{\bf k}_\parallel + R_\perp k_\perp)} + h.c.\right].
\]
For $N=3$, ${\bf k}$ is a $3$ dimensional vector with 
components ${\bf k_\parallel}$ and $k_\perp$ parallel and
perpendicular to the lattice plane.
${\bf R_\parallel}$ are chosen from the 
3 nearest neighbor lattice vectors 
for the honeycomb lattice,  and $R_\perp$ is
the perpendicular displacement between the two sublattices.  
$\tau^\pm$ guarantee that the hopping alternates between
the two lattice planes.  
This model is identical to the multi channel Kondo problem, 
with $t = J_{\perp}/2$,
provided the lattice constants are chosen to give the
appropriate scaling for $J_\perp$.
For $J_z = 0$, the dimension of $s^+_a(0)$ is 1, so
the cubic lattice constant in Fig. 2 is unity.  It follows
that $|{\bf R}_\parallel |^2 = 1-1/N$.  If the lattice
constant of the close packed Bravais lattice is $1/\sqrt{g}$,
then $g=1/2$. 
Finite $J_z$ may be treated nonperturbatively using
bosonization\cite{emery}, and affects the dimension of $s^+_a(0)$.
This leads to a distortion the lattice in the perpendicular
direction, 
$R_\perp^2 = (1- N J_z/2)^2/N$. 
Note that $R_\perp = 0$ for $J_z = 2/N$, so that the 
perpendicular direction decouples. 
This is the $N$ channel generalization of the Toulouse limit
\cite{emery,toulouse,Fabrizio}.
A central point of this paper is that this limit of the
N channel Kondo model is identical to the $g=1/2$ QBM
model on a $N-1$ dimensional ``honeycomb" lattice.

Note that the motion perpendicular to the planes alternates,
whereas the motion parallel to the planes does
not.  This gives rise to a renormalization of 
$R_\perp$ (or equivalently $J_z$) but not ${\bf R}_\parallel$.  
This may be seen from a RG analysis
similar to that of Anderson, Yuval, and Hamann \cite{anderson,kane3}.
Expressing the flow equations in terms of $J_z$ and $J_\perp$,
we find to order $J_\perp^3$, 
\begin{eqnarray}
dJ_z/d\ell &=& J_\perp^2 \left[ 1 - (N/2) J_z \right],\\
dJ_\perp/d\ell &=& J_\perp J_z \left[ 1 - (N/4) J_z \right]
 - (N/4) J_\perp^3.
\end{eqnarray}
The RG flows are shown in Fig. 3.  $J_z$ flows towards $2/N$,
the Toulouse limit, shown by the dashed line.  
For $N=3$ this is the same as the dashed line in Fig. 1b.  
The intermediate fixed point for the honeycomb lattice 
with $g=1/2$ is the same as the fixed point
of the 3 channel Kondo problem.
Varying $g$ adiabatically connects the multi channel Kondo fixed
point to the strong and weak barrier limits described perturbatively
in (7) and (9).  For large $N$, the fixed point 
at $J_\perp = J_z = 2/N$ approaches
the strong barrier limit, and is perturbatively accessible
\cite{Nozieres}.
Perturbations which break the symmetry between the two
sublattices are relevant and act like a magnetic field 
in the Kondo problem. The system then flows to a fixed 
point of the lower symmetry lattice.

Conformal field theory allows for 
an exact description of the multi channel Kondo fixed point\cite{ludwig}.
We can thus identify the critical exponents and mobility
for the $N-1$ dimensional generalized honeycomb lattice model for $g=1/2$.
The RG eigenvalue of the leading irrelevant operator at the fixed point
is $-2/(N+2)$.  The mobility is computed by 
identifying the appropriate correlation
function in the Kondo model.  The analogue of  ${\bf r}$ in the
Kondo model is the spin in each channel, $S_a^z$, projected onto the 
$N-1$ dimensional lattice plane with $\sum_a S_a^z$ constant.
This corresponds to the operator
$\hat O = \psi_{as}^\dagger \sigma^z_{ss'} T^A_{aa'} \psi_{a's'}$,
where $T^A$ is one of the $N-1$ diagonal generators of $SU(N)$.
The ``current" $\dot {\bf r}$ corresponds to a flow 
of spin between the different channels. 
Ludwig and Affleck\cite{ludwig2} have computed all correlation functions
of $\hat O$ exactly.  Borrowing their results, we obtain 
for the fixed point mobility,
\begin{equation}
\mu^* = 2 \sin^2 \frac{\pi}{N+2}.
\end{equation}
For $N=2$, the Kondo fixed point has $\mu^* = 1$, and is
at the ``small barrier" limit.  
For $N=3$, this value is plotted in Fig. 1b.  For
large $N$, the mobility and RG eigenvalue
may be found perturbatively
in a manner analogous to the $\epsilon$ expansion
following Eq. (9).  We have checked that they 
agree with the exact result to leading and sub leading
order in $1/N$.

\begin{figure}
\epsfxsize=2.8in
\centerline{\epsffile{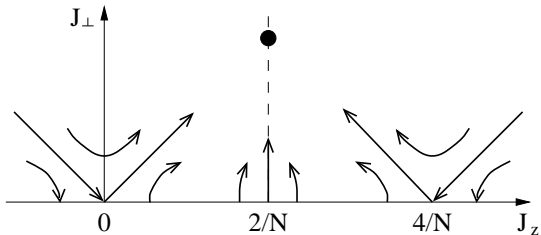}}
\caption{Flow diagram for the $N$ channel Kondo model for
small $J_\perp$.  The dashed line is the Toulouse
limit, $J_z=2/N$.  The strong coupling fixed point is marked
with the full circle.  }
\label{fig:kondo}
\end{figure}

Recently, Furusaki and Matveev\cite{furusaki} have studied
Coulomb blockade resonances in a spin degenerate quantum dot
with quantum point contact leads.
For $\Delta\ll T\ll E_C$, ($\Delta$ is the dot's level spacing and
$E_C$ is the Coulomb charging energy),
they argue that resonances are controlled by the 4 channel
Kondo fixed point.  This mapping may be understood 
in terms of the lattice of allowed charge states for
the dot and the four lead/spin channels:  a symmetric
dot on resonance has the symmetry of a diamond lattice with $g=1/2$.
Our analysis allows us to identify the
universal on  resonance conductance.  
A voltage between the leads corresponds to a force
$F = eV$, and the resulting current is 
$I =e |\dot{\bf r}|$.  From (12), with $N=4$,
$\mu^*=1/2$, leading to $G^* = (1/2)e^2/h$\cite{perfect}.

Our analysis also applies to resonant tunneling through a 
{\it single} resonant state (i.e. $T<\Delta$), for a multi channel 
Luttinger liquid with repulsive interactions\cite{kane2}.  
For two channels, a tunneling barrier maps to a two dimensional 
lattice with $g<1$, which, when tuned to a resonance, can in principle 
have a non symmorphic 
distorted honeycomb lattice symmetry.   Such resonances would be 
an adiabatic cousin of the 3 channel Kondo effect.  Analogous 
resonances have been observed in a single channel Luttinger 
liquid\cite{Webb}.

In summary, we have presented a general theory of quantum Brownian
motion on periodic lattices. For the honeycomb lattice and its
$D$ dimensional generalization, there is a non trivial intermediate
phase, which we have identified with the multi channel Kondo problem. 
Presumably, other non symmorphic lattices also display such phases,
and it would be interesting to classify them using 
conformal field theory.

We thank M.P.A. Fisher, K.A. Matveev and E.J. Mele for useful discussions
and comments.  This work has been supported by NSF grant DMR 95-05425.


\begin{references}

\bibitem{caldeira}
See for example: A.O. Caldeira and A.J. Leggett, Ann.\ Phys.\ (N.Y.)
 {\bf 149}, 374 (1983).

\bibitem{fisher}
\label{ref:fisher}
A. Schmid,  Phys.\ Rev.\ Lett.\ {\bf \bf 51}, 1506 (1983); 
M.P.A. Fisher and W. Zwerger, Phys.\ Rev.\ B {\bf 32},
 6190 (1985).

\bibitem{kane}
C.L. Kane and M.P.A. Fisher, Phys.\ Rev.\ Lett.\ {\bf 68}, 1221 (1992).

\bibitem{ludwig}
A.W.W. Ludwig and I. Affleck, Nucl.\ Phys.\ B {\bf 360}, 641
(1991).

\bibitem{Tarucha}
S. Tarucha, et. al., Solid State Com., {\bf 94}, 413 (1995).

\bibitem{Webb}
F.P. Milliken, et. al.,
Solid State Com. {\bf 97}, 309 (1996).

\bibitem{kane2}
C.L. Kane and M.P.A. Fisher, Phys. Rev. B {\bf 46}, 15233 (1992).

\bibitem{callan}
C.G. Callan, et. al., Nucl. Phys. B {\bf 442}, 444 (1995).

\bibitem{furusaki}
A. Furusaki and K.A. Matveev, Phys. Rev. B {\bf 52}, 16676 (1995).

\bibitem{henley} J. Kondev and C.L. Henley, 
preprint cond-mat/9511102.

\bibitem{andreas}
A.W.W. Ludwig, Int. J. Mod. Phys. B {\bf 8}, 347 (1994).

\bibitem{emery}
V.J. Emery and S. Kivelson, Phys.\ Rev.\ B {\bf 46}, 10812 (1992).

\bibitem{toulouse}
G. Toulouse, Phys.\ Rev.\ B {\bf 2}, 270 (1970).

\bibitem{Fabrizio}
M. Fabrizio and A.O. Gogolin, Phys. Rev. B {\bf 50}, 17732 (1994);
M. Fabrizio, A.O. Gogolin and P. Nozieres, Phys. Rev. B {\bf 51}, 16088 (1995).

\bibitem{anderson}
P.W. Anderson, G. Yuval, and D.R. Hamann, Phys.\ Rev.\ B {\bf 1},
 4464 (1970).

\bibitem{kane3}
A similar analysis for $N=2$ was performed in Ref. \onlinecite{kane2}.

\bibitem{Nozieres}
P. Nozi\`eres and A. Blandin, J. Phys. (Paris) {\bf 41}, 193 (1980).

\bibitem{ludwig2}
A.W.W. Ludwig and I. Affleck, Nucl.\ Phys.\ B {\bf 428}, 545
 (1994); Phys.\ Rev.\ B {\bf 48}, 7297 (1993).

\bibitem{perfect}
The ``perfect" conductance in this case is $e^2/h$ -- the series 
conductance of two incoherent $2e^2/h$ point contacts.
We thank K.A. Matveev for pointing this out.

\end{references}
\end{document}